# Placement of UAV-Mounted Mobile Base Station through User Load-Feature K-means Clustering

Amir Mirzaeinia[1], Mehdi Mirzaeinia[2], Mohammad Shekaramiz[3], Mostafa Hassanalian[4]

**Temporary high traffic requests in cellular networks is a challenging problem to address. Recent advances in Unmanned Aerial Vehicles applied to cover these types of traffics. UAV -Mounted Mobile Base Stations placement is a challenging problem to achieve high performance. Different approaches have been proposed; however, user required traffic is not considered in UAV placement. We propose a new feature to apply to K-means clustering to find the optimum clusters. User required traffic is defined as a new feature to assign users to the UAVs. Our simulation results show that UAVs could be placed closer to the high traffic users to achieve higher performance.**

## I. Introduction

Drones have a variety of applications in our daily life that have attracted the attention of many researchers around the world[1-6]. Therefore, everyday novel concepts of the drones that are able to fly and perform various missions autonomously in different environments are being developed[7-9]. Over recent years, drones or Unmanned Aerial Vehicles (UAVs) are increasingly applied in assisting 5G cellular networks to extend the coverage and service quality[10]. Especially, UAV assisted networks can be deployed on areas where extending infrastructure networks might not be possible or very expensive to develop[11]. Moreover, there are areas where there is a temporary traffic burst. These overcrowded events, such as football games, public demonstrations, and political protests, put a sudden extremely high demand for communication capacity on cellular networks around the duration of the events[12]. In these locations, there are not consistent high traffic requests, therefore it is not economical to deploy costly permanent infrastructures. Thus some agile on-demand mobile systems are required to cover these temporary service requests.

UAV systems have applied to a plethora of this type of applications over recent years[1-6]. Applying swarming UAV-mounted mobile base stations as on-demand cellular network infrastructure has been proposed to cover temporary high traffic requests. For example, Lyu et al.[13], in 2016, minimized the number of UAV-mounted mobile base stations (MBSs) needed to provide wireless coverage for a group of distributed ground terminals, ensuring that each terminal is within the communication range of at least one MBS. In this work, a polynomial-time algorithm was proposed for a successive MBS placement[13]. In 2016, Galkin et al.[14] applied a K-means clustering algorithm to partition the ground terminals to be served by p drones. Through numerical analysis, they demonstrated that clustering algorithms could be employed to position the aerial access points and select users to offload from the macrocells[14]. These systems can be programmed to move to any place to serve the users in a very short period of time. Fig. 1. shows a sample

---

[1] PhD Candidate, Department of Computer Science and Engineering, New Mexico Tech, Socorro, NM 87801, USA.

[2] Master Student, Department of Electrical Engineering, Amir Kabir University of Technology, Tehran, Iran.

[3] Assistant Professor, Electrical and Computer Engineering, Utah Valley university, Orem, Utah, 84058, USA.

[4] Assistant Professor, Department of Mechanical Engineering, New Mexico Tech, Socorro, NM 87801, USA

cellular network served by the UAVs-mounted mobile base station, which are connected to the core networks through satellite connections. In this paper, it is assumed that there is no base station tower infrastructure to cover the area and UAV systems are the only transmitters to serve the users in that area.

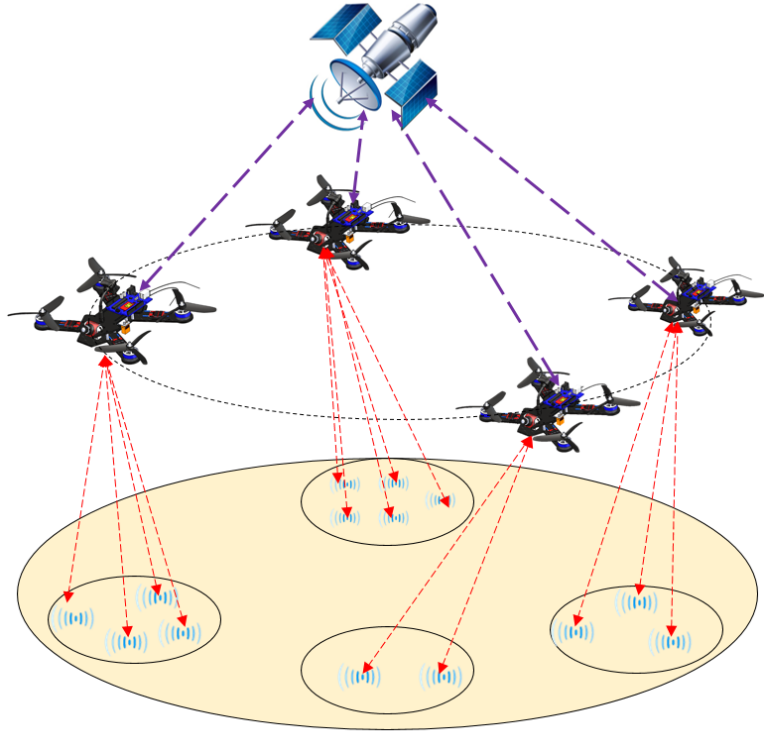

Figure 1. Cellular networks served by UAV-Mounted Mobile Base Station.

In spite of all advantages of the UAV-Mounted Mobile Base Station, there are some challenges that need to be addressed. UAV-Mounted Mobile Base Stations placement is one of the main challenges that researchers are investigating to find the optimum values. Although different methods have already proposed to find the proper places of the UAVs, to the best of our knowledge, they only consider the number of users as the traffic load, which is not a realistic scenario. In more realistic scenarios, different users usually run different applications, such as web browsing, voice call or video calls. Therefore, users consume and require a variety of radio link bandwidth and quality. In this paper, a weighted K-means clustering approach is used to find the cluster centroid.

## II. UAV systems and Swarm

UAVs can be designed very differently, which highly depends on the variety of the mission requirements. Drones can be classified based upon their characteristics, such as size, weight, flight endurance and capabilities. Hovering flight mode is one of the major capabilities of the UAVs that is essential in many civil applications. Hovering capability is yet another factor in classifying the unmanned aerial systems. Rotary wings, multi-rotors, tilt-rotors, and tilt-wings are some type of drones that can hover while they perform their missions. In this current application, UAV-Mounted Mobile Base Station, a UAV system with hovering capability should be used to serve radio network users. Navigation is also another major part of the UAV systems. A manual remote control is the very first and simple control system to navigate drones to their final mission location. However, recent advances in autonomous systems help drones to navigate automatically based on their neighbors and environments. This type of autonomous systems can deploy GPS information to find their final positions.

## III. Pre-Processing and Clustering Approach

User to UAV assignment can be thought of as a clustering task that could be performed based on various parameters. There are types of clustering algorithms proposed in different fields and applications. Various algorithms differ significantly in their understanding of what constitutes a cluster. The most popular type of clustering algorithm is grouping the objects based on their distances. Distance definition can differ in various applications. Connectivity-based, centroid-based, distribution-based, density-based, and grid-based clusterings are different types of clustering that are applied to a variety of applications. K-means is one of the most popular centroid based clustering algorithms that is applied in this study for users to UAV assignment application.

K-means clustering approach is one of the well-known clustering methods[15-17]. Using this method, each user will be served by the cluster with the nearest UAV-Mounted Mobile Base Stations. K-means is the iterative clustering approach to find the best centroid, which is the position of the UAV in this problem. K-means algorithm initiates by random positioning of the UAVs and iteratively updates the location of the centroid and cluster borders. Considering the user traffic weights, an idea of considering user traffic load as another feature for clustering purposes is proposed. For example, a user who is consuming one-megabyte download traffic can be considered as a user with a load feature of one. A typical user to UAV assignments, *x* and *y* location of the users are considered as the main features. In this scenario, user traffic load as the third feature to the user *x* and *y* location is added. The following algorithm shows the K-means clustering steps.

---

**Algorithm 1. K-means Clustering Algorithm**

---

**Preprocessing**: Each user split to a number of users with minimum possible traffic.

**Initialization:** Set K-means UAVs to random positions:

**Assignment step:** Each user assigned to the nearest UAV. Euclidean distance is used to find the distance between users and cluster centroids.

$$s_i^t = \{x_p : \left| x_p - m_i^t \right|^2 \leq \left| x_p - m_j^t \right|^2 \forall j, 1 \leq j \leq k\}$$

where in this equation *x* is the set of users, *m* is the set of means of clusters, *k* is the number of clusters, and *t* is the iteration number.

**Update centroid:** recalculating the centroid for the assigned users.

$$m_i^{t+1} = \frac{1}{\left| s_i^t \right|} \sum_{x_j \in s_i^t} x_j$$

Repeating these steps will converge to the step that assignments no longer change.

---

K-means clustering uses features to assign users to the UAVs. In this proposed method, user traffic as another feature of the system is added to find user clusters and compare the results with the two-dimensional geographic location features.

## IV.         Experiments and simulation results

In this experiment, random users distributed around the area with two levels of required traffics are generated. First, the K-means that only works based on the 2-d Euclidean positional distance is applied, which is essentaialy based on two coorniate features. In the the second scenario, the required traffic is employed as the third feature. Fig. 2. (a) shows two-features user assignments and Fig. 2(b) illustrates the traffic applied to three-features user assignments. As it is demonstrated in this figure, there are three users with high traffic required which are located in the border of the two-feature clustering approach (Fig.2.(a)) while these three users are located in a high-quality region of the three-features clustering approach (Fig. 2.(b)). As demonstrated in Fig. 2 (a)., all users are considered as the same load users. Then, they are presented as the same size orange dots. However, in Fig.2 (b), high-speed users are presented as dots larger than the low speed required users. Considering the dynamic behaviour of the users, UAV adaptive placement is another problem that needs to be addressed. Therefore, in the future direction, the clustering processing time will be assessed to minimize realtime user behavior. Moreover, the maneuvering capability of UAV systems helps to change the coverage radius of the cells to adapt to traffic changes. Analyzing the power change or altitude of the UAV system is another future work to find the minimum required power to adapt to traffic changes.

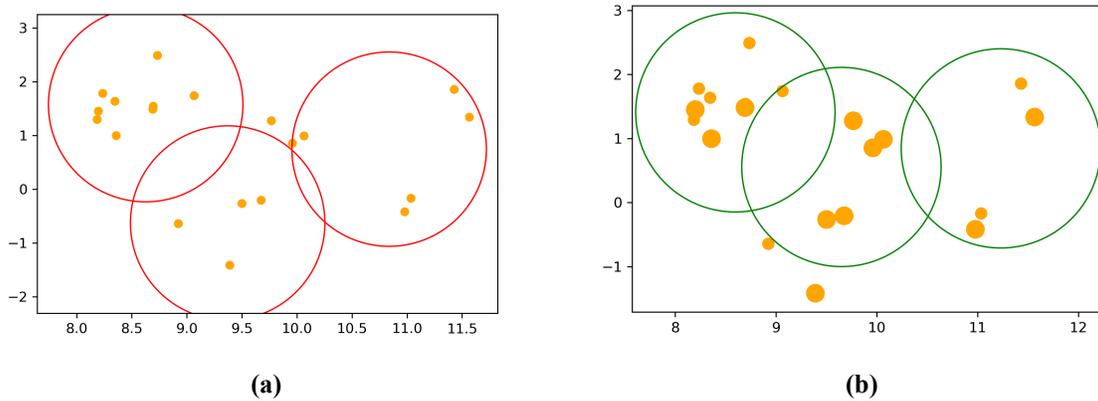

(a)                                              (b)

**Figure 2. Views of (a) two-feature K-means clustering and (b) three-features K-means clustering.**

## V. Conclusions

UAV-Mounted Mobile Base Station is a new approach to cover places where there exist temporary high traffic requests. UAV placement is a challenging problem to achieve the optimum service. Crowd location-based UAV placement is proposed to cover the crowd's sudden high requests. However, this approach still needs to be investigated and modified to meet the design of more advanced networks such as 4G and 5G because they can provide high-speed links. The method proposed method in this paper considers the user traffic as an additional user feature to cluster. Therefore, UAV can be placed closer to high speed required users to be able to serve more.